\documentclass[aps,pra,twocolumn,nofootinbib]{revtex4-1}
\usepackage{epsfig}
\usepackage[colorlinks,linkcolor=blue,anchorcolor=blue,citecolor=blue,urlcolor=blue,breaklinks=true]{hyperref}
\usepackage{graphics}
\usepackage{color}
\usepackage{amsmath}
\usepackage{bm}
\begin{document}
\author{Guo-Hua Liang$^{1}$}
\email{lianggh@njupt.edu.cn}
\author{Meng-Yun Lai$^{2}$}

\address{$^{1}$ School of Science, Nanjing University of Posts and Telecommunications, Nanjing 210023, China}
\address{$^{2}$ College of Physics and Communication Electronics, Jiangxi Normal University, Nanchang 330022, China}

\title{Effective quantum dynamics in curved thin-layer system with inhomogeneous confinement}
\begin{abstract}
The motion of quantum particles homogeneously constrained to a curved surface is affected by a curvature induced geometric potential. Here, we consider the case of inhomogeneous confinement and derive the effective Hamiltonian by extending thin-layer procedure, where an extra effective potential appears. This effective potential is relevant to the ground state energy perpendicular to the surface and the morphology of the confining potential. Tiny changes in the thickness are envisioned to induce considerable magnitude of the effective potential. To demonstrate the impact of the inhomogeneity, we apply our method to investigate the coherent transport on a cylindrical surface where a confining potential with two helical ditches is imposed. Numerical analysis reveals that the inhomogeneity of the confinement significantly changes the transport properties. This study develops the method for low-dimensional constrained systems and exhibits the possibility of new degree of control for waveguiding in nanostructures.
\end{abstract}


\maketitle
\section{INTRODUCTION}
With the constant progress of techniques for the synthesis of nanostructures~\cite{ph500144s,C4NR00330F,201705630,Pogosov_2022}, much attention is paid to the corresponding descriptions of various dynamics in low-dimensional systems, in which different kinds of geometric quantities play important roles. Among these quantities, curvature appeals to many interests since it significantly breaks symmetries associated with space coordinates, and often relates the system with general relativity, offering a glimpse on the effect of strong gravitational field. For example, two beam interference~\cite{PhysRevLett.105.143901}, the evolution of speckle patterns~\cite{RN184}, the phase and group velocities of wavepackets~\cite{Bekenstein2017Control} and elastic waves~\cite{PhysRevLett.121.234301} on curved surfaces have been investigated experimentally, providing analogue models for wave optics in the gravitational fields of black holes, wormholes and universe with non-vanishing cosmological constant. Besides, the study of curvature effects covers more and more areas nowadays, such as biological systems~\cite{2005Membrane,RN1} and pattern dynamics~\cite{PhysRevLett.128.224101,RN2,1.5108838,SAITO2021103087}, semiconductors~\cite{PhysRevB.84.045438,PhysRevB.86.195421,Chang_2017}, magnetism~\cite{1.4941045,adma.202101758}, superconductivity~\cite{PhysRevB.79.134516,PhysRevB.94.081406} and phase transition~\cite{D0SM00652A}, manifesting novel and unfamiliar phenomena compared with planar cases.

For the theoretical description of motions bounded to an arbitrary curved surface, there is a generic procedure to obtain the two dimensional (2D) effective Hamiltonian or equation called thin-layer procedure (TLP) or confining potential approach. This approach is originally introduced for tackling the limiting situation where a quantum particle in three dimensional (3D) Euclidean space is constrained to a curved surface by a strong confining force~\cite{JENSEN1971586,PhysRevA.23.1982}. It was found a geometric potential depending on the intrinsic and extrinsic curvature of the surface appears in the effective equation, which was demonstrated in photonic crystals later~\cite{PhysRevLett.104.150403}. The geometric potential reveals that even the dynamics in the normal direction is "frozen" by the confining potential, it indeed contributes to the tangential dynamics because of the curvature. In light of this, TLP has been developed and applied to many situations, such as a charged particle in an electric and magnetic field~\cite{PhysRevLett.100.230403,Brandt_2015,Wang2016a}, Dirac particles~\cite{OUYANG1999297,PhysRevA.48.1861,BRANDT20163036}, spin-orbital coupling~\cite{PhysRevB.91.245412,PhysRevB.64.085330,PhysRevB.87.174413,PhysRevA.90.042117,Wang2017,PhysRevA.98.062112} and an electromagnetic field~\cite{PhysRevA.78.043821,PhysRevA.97.033843,PhysRevA.100.033825}.
In these studies, in addition to the scalar geometric potential, the curved features lead to more geometric effects which associated with the internal degree of freedom and property of the confined particles. TLP was also extended to the case that an arbitrary manifold embedded in a higher-dimensional Euclidean space, it is found that a gauge potential appears in the effective Hamiltonian when the space of states for the direction normal to the surface is degenerate~\cite{S0217732393000891,Maraner_1995,MARANER1996325,S0217751X97002814,SCHUSTER2003132}.

It is noteworthy that, most studies on the thin layer system considered the case where the confining potential is homogeneous everywhere around the surface.
In this work, by generalizing TLP, we discuss the dynamics of a quantum particle constrained to an arbitrarily curved surface by an inhomogeneous confining potential. This problem is important for two reasons: one is to estimate the influence from the imperfection of the confinement, since it is difficult to ensure the thickness of the layer is the same everywhere in reality; the other is to offer a new toolbox to manipulate and guide quantum states by designing the feature of the confining potential. Besides the curvature, the confinement inhomogeneity could bring more possibilities to tune the properties of nanostructures. In optics, the variable-thickness of a microstructured waveguide can induce an effective refractive index, which is extracted from experimental data fitting~\cite{RN3}. In this work, we will analytically show that in quantum mechanics the inhomogeneous confining potential can induce an effective potential totally different from the curvature induced geometric potential.

The structure of this paper is as follows. In section~\ref{s2}, we derive the effective Hamiltonian for particles constrained to an arbitrary curved surface by an inhomogeneous confinement. In section~\ref{s3}, we apply this formalism in the case of a cylinder with extra helical confinement force acting on the surface and numerically investigate the transport properties. In section~\ref{s4}, we summarize the conclusions.
\section{Effective dynamics}\label{s2}
In 3D Euclidean space, the geometry of a curved surface $\mathcal{S}$ can be described by a position vector $\bm{r}(q_1,q_2)$, where $(q_1,q_2)$ the curvilinear coordinates. In order to study the quantum mechanics of a spinless particle confined to $\mathcal{S}$, we need describe the portion of space in an immediate neighbourhood of $\mathcal{S}$. Conventionally in TLP, adapted coordinates $(q_1,q_2,q_3)$ are always chosen to parameterize the space as

\begin{equation}
\bm{R}(q_1,q_2,q_3)=\bm{r}(q_1,q_2)+q_3 \bm{N}(q_1,q_2),
\end{equation}
where $\bm{N}(q_1,q_2)$ is the unit vector normal to the surface, and $|q_3|$ gives the distance from the surface. Thus when the distribution of the confining potential $V_c$ is homogeneous on the surface, it depends only on the normal coordinate, $V_c=V_c(q_3)$. Usually, we assume that $V_c$ has a deep minimum at $q_3=0$ and is symmetric in the normal direction about its minimum. Therefore, we can expand $V_c$ as a power series in the $q_3$,
\begin{equation}\label{vcq}
V_c(q_3)=\frac{m}{2}\omega^2 q_3^2+O[(q_3)^3],
\end{equation}
where $m$ is the particle mass, and the frequency $\omega $ is an intensity parameter.

Now let us consider the inhomogeneous confinement case.
Here, we suppose that the confining potential can be written as $U_c(q_1,q_2,q_3)=s^2(q_1,q_2) V_c(q_3)$, where $s(q_1,q_2)$ is a dimensionless and continuous function close to unity, which determines the morphology of the potential on $\mathcal{S}$. This expression indicates that $U_c$ is also symmetric about the minimum $q_3=0$, although it varies with $q_1$ and $q_2$.

Considering the inhomogeneity of the confining potential, we parameterize the neighbourhood space of the surface in new coordinates $(q_1,q_2,Q_3)$, where $Q_3=\bar{s}(q_1,q_2)q_3$ with $\bar{s}(q_1,q_2)$ is an undetermined function. The new parametrization is then
\begin{equation}\label{nrr}
\bm{R}(q_1,q_2,Q_3)=\bm{r}(q_1,q_2)+\frac{Q_3}{\bar{s}(q_1,q_2)}\bm{N}(q_1,q_2).
\end{equation}
In this coordinates, the irregularities of the confining potential are expected to be absorbed in the normal coordinate $Q_3$. Applying this parametrization, we can calculate
the covariant components of the three-dimensional metric tensor via $G_{ij}=\partial_i \bm{R}\cdot \partial_j \bm{R} $ with $i,j=1,2,3$.
From Eq.~\eqref{nrr} we obtain
\begin{equation}\label{pa3}
\begin{aligned}
\partial_a \bm{R}&=\partial_a \bm{r}+\frac{ Q_3}{\bar{s}}[\partial_a\bm{N}(q_1,q_2)]+\left[ \partial_a \left(\frac{1}{\bar{s}} \right) \right]Q_3 \bm{N}(q_1,q_2), \\
\partial_3 \bm{R}&=\frac{1}{\bar{s}}\bm{N}(q_1,q_2),
\end{aligned}
\end{equation}
where the index $a=1,2$ (so does $b,c$ and $d$ in the text below).
Since the derivatives of the normal vector $\hat{N}(q_1,q_2)$ lie in the tangent plane of the surface, we have
\begin{equation}
\partial_a\bm{N}=\alpha_{ab}\partial_b \bm{r},
\end{equation}
where $\alpha_{ab}$ is called Weingarten curvature matrix.
Thus, we obtain all components of the metric tensor $G_{ij}$,
\begin{equation}
\begin{aligned}
G_{ab}=\gamma_{ab}+Q_3^2\left[\partial_a\left( \frac{1}{\bar{s}}\right)\right] \left[\partial_b \left(\frac{1}{\bar{s}}\right)\right]
\end{aligned}
\end{equation}
and
\begin{equation}
G_{a3}=G_{3a}=\frac{1}{\bar{s}}\left[\partial_a\left( \frac{1}{\bar{s}}\right)\right]Q_3, \ \ G_{33}=\frac{1}{\bar{s}^2},
\end{equation}
where
\begin{equation}
\gamma_{ab}=g_{ab}+\frac{Q_3}{\bar{s}}[\alpha g+(\alpha g)^T]_{ab}+ \frac{Q_3^2}{\bar{s}^2} (\alpha g \alpha)_{ab}
\end{equation}
and $g_{ab}=\partial_a \bm{r}\cdot \partial_b \bm{r}$ is the 2D metric tensor for the surface $\mathcal{S}$.
The determinant of $G_{ij}$ can also be worked out and the result is $G=|\gamma|/\bar{s}^2$.

Further calculation gives the exact form of the inverse of the metric tensor, which turns out to be
\begin{equation}\label{gin}
G^{ij}=\left(
\begin{array}{ccc}
\lambda^{ab} & \lambda^{ac}Q_3 (\partial_c \bar{s})/\bar{s} \\
\lambda^{bc}Q_3 (\partial_c \bar{s})/\bar{s} & \bar{s}^2+Q_3^2 (\partial_c \bar{s}) \lambda^{cd}(\partial_d \bar{s})/\bar{s}^2
\end{array}
\right),
\end{equation}
where $\lambda^{ab}=(\gamma_{ab})^{-1}$ is the inverse of $\gamma_{ab}$.

We can now turn our attention to the derivation of the effective Hamiltonian. The 3D Hamiltonian containing the confining potential $U_c$ can be written in the curvilinear coordinates $(q_1,q_2,Q_3)$ as
\begin{equation}\label{seq}
H_{\text{3D}}=-\frac{\hbar^2}{2m}\nabla^2+s^2V_c(Q_3/\bar{s}).
\end{equation}
From Eq.~\eqref{gin}, the explicit form of the Laplacian is
\begin{equation}
\begin{aligned}
\nabla^2=&\frac{1}{\sqrt{G}}\partial_i \sqrt{G}G^{ij}\partial_j \\
=&\frac{1}{\sqrt{G}}\partial_3 \sqrt{G}[\bar{s}^2+Q_3^2(\partial_c \bar{s}) \lambda^{cd}(\partial_d \bar{s})]\partial_3 \\
&+\frac{1}{\sqrt{G}}\partial_a \sqrt{G}\lambda^{ab}\partial_b
+\frac{1}{\sqrt{G}}\partial_a\sqrt{G}\lambda^{ac}Q_3 \frac{\partial_c \bar{s}}{\bar{s}}\partial_3 \\ &+\frac{1}{\sqrt{G}}\partial_3\sqrt{G}\lambda^{bc}Q_3 \frac{\partial_c \bar{s}}{\bar{s}}\partial_b.
\end{aligned}
\end{equation}

The corresponding wavefunction $\Phi$ satisfies the normalization condition
\begin{equation}
\int |\Phi|^2 \sqrt{G}dq_1 dq_2 dQ_3=1.
\end{equation}
Our purpose is to get an effective 2D Hamiltonian whose wavefunction describes the quantum probability density on the surface $S$. Therefore, we need rescale the 3D wavefunction $\Phi$ by $(|G|/|g|)^{1/4}$, namely $\Psi=(|G|/|g|)^{1/4}\Phi$, where $g$ is the determinant of $g_{ab}$. The normalization of the new wavefunction $\Psi$ is then
\begin{equation}
\int |\Psi|^2 dQ_3\sqrt{g}dq_1 dq_2 =1.
\end{equation}
According to this condition, one can regard $\int|\Psi|^2 dQ_3$ as a probability density for a particle moving on $\mathcal{S}$ with the curvilinear measure $\sqrt{g}dq_1 dq_2$.
Accordingly, the Hamiltonian should also be rescaled as $H=(|G|/|g|)^{1/4}H_{\text{3D}}(|G|/|g|)^{-1/4}$.
By introducing operators $\hat{\partial}_a=\partial_a+iA_a L_3$, where $A_a=\frac{\partial_a \bar{s}}{\bar{s}}$, and $L_3=-iQ_3\partial_3$, the rescaled Hamiltonian can be written in a compact form,
\begin{equation}\label{hrs}
\begin{aligned}
H=&-\frac{\hbar^2}{2m}\left[\bar{s}^2 G^{-\frac{1}{4}} \partial_3\sqrt{G}\partial_3 G^{-\frac{1}{4}} \right. \\
& \left. +(gG)^{-\frac{1}{4}}\hat{\partial}_a\sqrt{G}\lambda^{ab}\hat{\partial}_b (g/G)^{\frac{1}{4}}\right]+s^2V_c(Q_3/\bar{s}).
\end{aligned}
\end{equation}

Up to now, no approximation has been made. In order to get the effective Hamiltonian, we need separate the dynamics perpendicular and tangent to the surface, which is associated with the explicit form of the confining potential.
Here, we return to Eq.~\eqref{vcq} and neglect the terms of order $(q_3)^3$ and higher, which leads to a harmonic binding form.
As a confining potential, $U_c=\frac{s^2}{\bar{s}^2} \frac{m}{2}\omega^2 (Q_3)^2$, must has a large $\omega$ to ensure the quantum well is deep enough.
To evaluate the magnitude of $\omega$, following the approach of TLP, we introduce a small dimensionless parameter $\epsilon$ and rescale the harmonic frequency as $\omega\rightarrow \omega/\epsilon$.
Because of the binding, wavefunction will be squeezed in a very small range around $Q_3=0$ in the transverse direction.  Adopting $\epsilon$ as a perturbative parameter, we also rescale the normal coordinate as $Q_3\rightarrow \sqrt{\epsilon}Q_3$.
In this way, $U_c$ is of the order $\epsilon^{-1}$. The Hamiltonian can be written in powers of $\epsilon$ as
\begin{equation}\label{h01}
H=H_0+H_1+O(\epsilon^{1/2})
\end{equation}
where
\begin{equation}
H_0=\frac{1}{\epsilon}\left[-\frac{\hbar^2}{2m}\bar{s}^2\partial_3^2+\frac{s^2}{\bar{s}^2} \frac{m}{2}\omega^2 (Q_3)^2 \right]
\end{equation}
and
\begin{equation}\label{h1}
H_1=-\frac{\hbar^2}{2m}\frac{1}{\sqrt{\bar{s}}}\left[\frac{1}{\sqrt{g}}\hat{\partial}_a \sqrt{g}g^{ab}\hat{\partial}_b \right]\sqrt{\bar{s}}+V_g.
\end{equation}
Here, $V_g=-\frac{\hbar^2}{2m}(M^2-K)$ is the well-known geometric potential, with the mean curvature $M=\text{Tr}(\alpha_{ab})/2$ and Gaussian curvature $K=\det(\alpha_{ab})$. During performing the limit, we have used the fact that $\sqrt{|\gamma|}=[1+\text{Tr}(\alpha_{ab})Q_3/\bar{s}+\det(\alpha_{ab})(Q_3)^2/\bar{s}^2]\sqrt{|g|}$.

From Eq.~\eqref{h01}, when $\epsilon\rightarrow 0$, only $H_0$ and $H_1 $ survive. $H_0$ is of the order $\epsilon^{-1}$, which describes a particle bounded by the confining potential in the transverse direction and takes a lead role in $H$. Being of the order $\epsilon^{0}$, $H_1$ corresponds to the quantum dynamics in the tangential direction on $\mathcal{S}$. However, we are aiming at investigating the tangential behavior on the surface in the energy range where the quantum particle is in the ground state in the transverse direction. Therefore, the effective Hamiltonian should be of the order $\epsilon^{0}$. To get the effective 2D Hamiltonian we need consider the Schrodinger equation $(H_0+H_1)\Psi=E\Psi$, where $E$ denotes the total energy, and the separation of the wavefunction.

Taking into account $\bar{s}$ is undetermined, we set $\bar{s}=\sqrt{s}$, and multiply $1/s$ on both sides of the equation. The equation can be rewritten as follows,
\begin{equation}\label{seq1}
\frac{1}{\epsilon}\left[-\frac{\hbar^2}{2m}\partial_3^2+ \frac{m}{2}\omega^2 (Q_3)^2 \right]\Psi+\frac{H_1}{s}\Psi=\frac{E}{s}\Psi.
\end{equation}
Formally this equation is easily separated. Making the assumption $\Psi=\psi(q_1,q_2)\chi(Q_3)$, the usual variable separation gives
\begin{equation}\label{hoe}
[-\frac{\hbar^2}{2m}\partial_3^2+ \frac{m}{2}\omega^2 (Q_3)^2 ]\chi=\epsilon E_0 \chi,
\end{equation}
and
\begin{equation}\label{eeq}
H_1\psi=(E-s E_0)\psi.
\end{equation}
It is clear that Eq.~\eqref{hoe} describes a one-dimensional harmonic oscillator, and the corresponding energy eigenvalue $E_0$, which is of the order $\epsilon^{-1}$, is the dominant part of the total energy $E$. Eq.~\eqref{eeq}, which describes the 2D effective dynamics on $\mathcal{S}$ under the transverse mode energy $E_0$, can be rewritten as
\begin{equation}
[H_1+(s-1)E_0]\psi=E_1\psi,
\end{equation}
where $E_1=E-E_0$. Here, it should be emphasized that the dynamics separation is based on the perturbative parameter $\epsilon$. To ensure the validity of the separation, and taking into account $E_0\sim \epsilon^{-1}$, we have to limit the function $s(q_1,q_2)$ that $(s-1)\sim \epsilon$. This limitation also implies the application range of the method. Up to now we are able to define the effective 2D Hamiltonian
\begin{equation}\label{hef}
H_{\text{eff}}=\int \chi^*[H_1+(s-1)E_0]\chi dQ_3.
\end{equation}
To calculate this integral, one need to utilize
\begin{equation}
\int \chi^* L_3 \chi dQ_3=\frac{i}{2}.
\end{equation}
Performing the integral in Eq.~\eqref{hef} we eventually obtain the explicit form of the effective Hamiltonian
\begin{equation}\label{heff}
H_{\text{eff}}=-\frac{\hbar^2}{2m}\frac{1}{\sqrt{g}}\partial_a\sqrt{g} g^{ab}\partial_b+V_g+(s-1)E_0.
\end{equation}
It should be noted, that no gauge potential appears in this effective Hamiltonian, despite the operator $\hat{\partial}_a$ in Eq.~\eqref{hrs} contains the corresponding terms. This is due to that the factor $\sqrt{\bar{s}}$ in Eq.~\eqref{h1} cancels out these terms after the integration, which is expectable as otherwise the Hamiltonian would be non-Hermitian. In contrast to the geometric potential $V_g$, the effective potential $(s-1)E_0$ stemming from the inhomogeneous confinement is relevant to the ground state energy in the normal direction of the surface. Typically, in a thin-layer system, the thickness $d$ is much smaller than the curvature radius $r_c$. If we set $\epsilon=d/r_c$, $s-1$ should be of the order $d/r_c$ at most, which implies that tiny changes of the layer thickness can induce considerable influences.
\section{coherent transport in a cylinder with inhomogeneous confinement}\label{s3}
\begin{figure}
  \centering
  \includegraphics[width=0.45\textwidth]{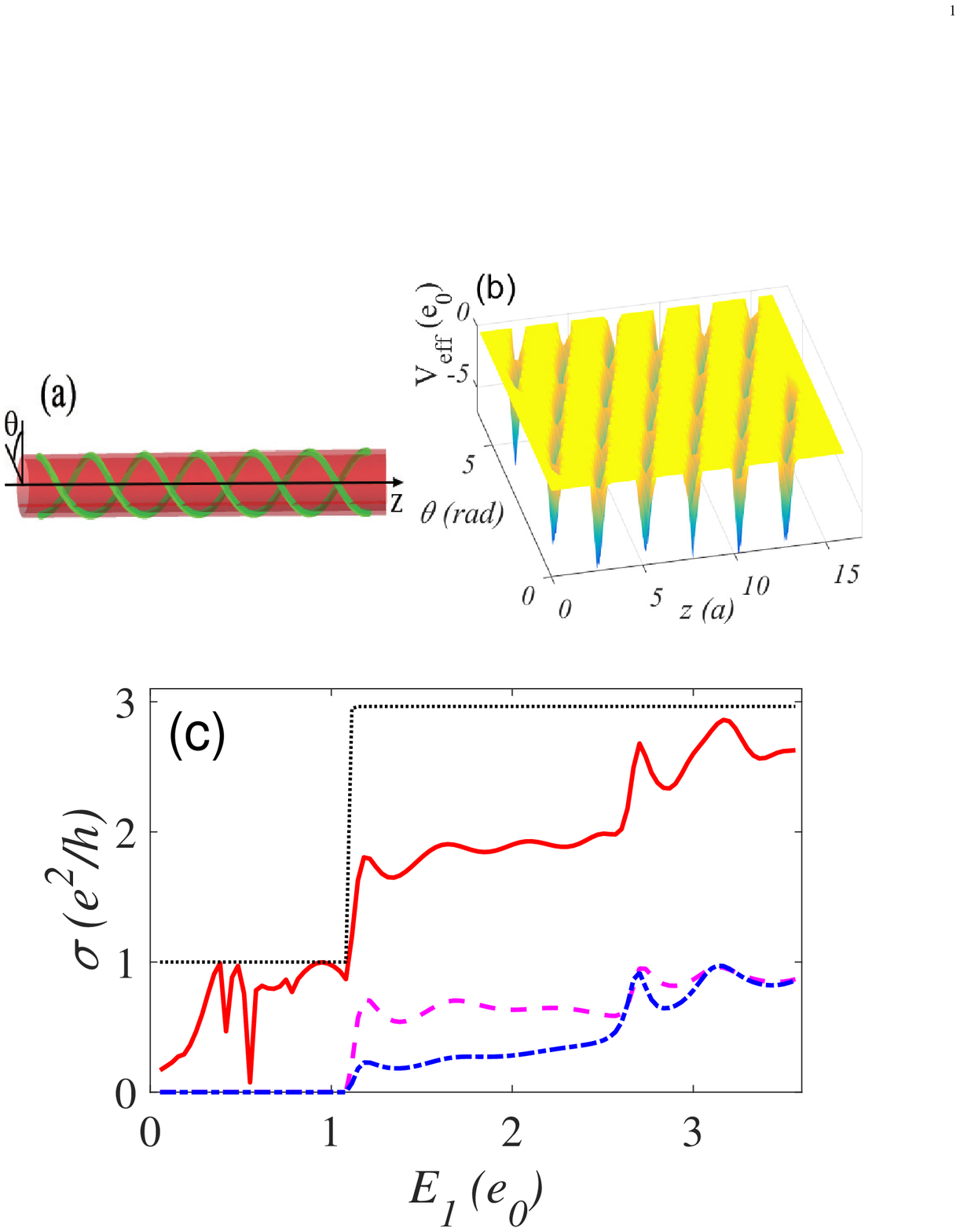}\\
  \caption{(a) Schematic of a cylindrical surface. Inhomogeneous confining potentials are arranged along the green helical curves. (b) Effective potential $E_{\text{eff}}$ in $(\theta,z)$ coordinates. (c) Conductance versus energy $E_1$. The black dotted line and red solid line are the total conductance for the cylindrical layer with homogeneous and inhomogeneous confinements, respectively. The pink dashed and blue dash-dotted line are for the case of injected mode $l=1$ and $l=-1$, respectively. }\label{fig1}
\end{figure}

In this section we give an example of a cylindrical surface with inhomogeneous confining potentials which has helical characters (see Fig.~\ref{fig1}(a)). As shown in the figure, we assume that two dimensional electron gases are confined to such a cylindrical surface with a radius $r$, and the confining potential is homogeneous except two ditches lie in the strip along the green helical lines. Such a structure may be realized by lithography techniques. According to the effective Hamiltonian we have derived, the inhomogeneous confinement induces an effective potential, which will impact the transport properties of the cylindrical tube.

We suppose the incident wave goes from the left to the right, and the wavefunctions of the injection states are $\psi_{in}=\frac{1}{\sqrt{2\pi}} e^{il\theta }e^{ikz} $, where $\theta $ is the azimuthal angle and $k=\sqrt{2m(E_1-E_l)}/\hbar$ with $E_l=\frac{l^2 \hbar^2}{2mr^2}$. Here, $E_1=E-E_0$ with $E_0$ the ground state energy of the quantum well normal to the cylinder in the homogeneous area ($s(\theta,z)=1$). For convenience, we scale the length and energy in units of $a$ and $e_0=\hbar^2/(2m a^2)$.

To realize a inhomogeneous confinement with a helical characteristics, we design the function $s$ to be
\begin{equation}
s(\theta,z)=1-\epsilon\left[\frac{1}{2}\cos \Omega(r\theta-\kappa z+n\pi)+\frac{1}{2}\right],
\end{equation}
where $n=0,\pm 1,\pm 2, \cdots$, $\kappa$ is a parameter to control the tilt angle of the green line in Fig.~\ref{fig1}(a) and $\Omega$ is a parameter to adjust the width of the strips. Such a function describes a slightly weaker confining potential or bigger thickness in the strips region. The induced effective potential $V_{\text{eff}}$ is described in Fig.~\ref{fig1}(b) in $(\theta ,z)$ coordinates for $E_0=70e_0$, $\epsilon=0.1$ and $\Omega=8a^{-1}$. By numerically solve the corresponding effective Schrodinger equation, we calculate the transmissions for the modes $l=0,\pm 1$. Based on Landauer formula~\cite{Landauer1987}, the conductance at zero temperature is obtained and plotted in Fig.~\ref{fig1}(c). We first plot the conductance (black dotted line) for the case that the confinement is homogeneous everywhere, which shows a perfect step-like dependence on $E_1$, and the height of the step from $1\sigma_0$ to $3\sigma_0$ ($\sigma_0=e^2/h$, with $e$ the electric charge and $h$ the Planck constant) indicates modes $l=\pm 1$ are degenerate in this situation. While for the inhomogeneous case (red solid line), we observe that the conductance is more complicated and a new plateau of $2\sigma_0$ from $1.2e_0$ to $2.5e_0$ is formed. This plateau shows that the inhomogeneity of the confining potential destroys the degeneration of modes $l=\pm 1$ and only one open channel appears in this energy range. To make it more clear, we draw the conductances for the injected modes $l=\pm 1$, where an evident difference between the two lines emerges, showing that the wave in mode $l=1$ is more preferable to pass through the cylindrical structure than the mode $l=-1$ in this energy range.

\begin{figure}
  \centering
  \includegraphics[width=0.45\textwidth]{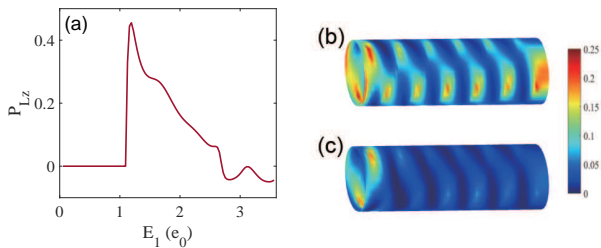}\\
  \caption{(a)Mean angular momentum polarization $P_{L_z}$ of the outgoing current at the right side. (b) and (c) The probability densities for transport when injected modes are $l=1$ and $l=-1$, respectively. $E_1=1.3e_0$. }\label{fig2}
\end{figure}

We further define a polarization quantity $P_{L_z}$ to show the ratio between mean angular momentum current to the total currrent, which is
\begin{equation}
P_{L_z}=\sum_{l'}\frac{\sigma_{l', l}-\sigma_{l', -l}}{\sigma},
\end{equation}
where $\sigma_{l',l}$ denote the conductance that incident modes $l'$ are scattered into modes $l$. In Fig.~\ref{fig2}(a) we plot the dependence of $P_{L_z}$ on energy $E_1$. It shows that once the threshold energy of modes $l=\pm 1$ is reached, angular momentum polarization is generated in the outgoing current and has a rapid increase up to a maximum value. With $E_1$ increasing, the polarization decrease slowly, which is due to the counteraction of the arising conductances $\sigma_{l',-1}$ carrying the opposite angular momentum. As the chirality possessed by the system, it is easy to predict that the propagation in the opposite direction (right to left) results in the same magnitude but opposite angular momentum polarization. To visually comprehend the generation of the polarization, the probability densities are plotted in Fig.~\ref{fig2}(b) and (c) at the energy $E_1=1.3e_0$ for the injected mode $l=1$ and $l=-1$, respectively. We can find that the injected wave in the mode $l=1$ is able to be transmitted to the right side, while the wave in the mode $l=-1$ is mostly reflected. It should also be noted that the probability density is prominently higher along the helical strips (the green lines in Fig.~\ref{fig1}(a)), indicating a helical and open channel is formed for the mode $l=1$. It demonstrates that by implementing such kinds of inhomogeneous confinement, waveguides may be able to be fabricated on curved surfaces or substrates, which are rarely explored.

\section{Conclusion}\label{s4}
In summary, we have extended TLP and derived an effective Hamiltonian for a particle constrained to an arbitrary curved surface by inhomogeneous confining potentials. Due to the inhomogeneity of the confinement, we find that an effective potential is induced, which is proportional to the ground state energy in the normal direction and is also determined by the feature of the confining potential. We apply the method to a cylindrical surface where the confining potential is designed to have two additional helical ditches, and numerically study the transport properties. It is shown that the helicity of the confinement destroys the degeneration of the excitation modes and leads to the generation of angular momentum polarization in the outgoing current. This method can serve as a tool to conveniently get the information of energy bands and transport properties for various low-dimensional nanostructures, and enable the designation and ``writting" of waveguides on arbitrary geometries.

\acknowledgments

This work is supported in part by the National Natural Science Foundation of China (under Grants No. 12104239), National Natural Science Foundation of Jiangsu Province of China (under Grant No. BK20210518), Nanjing University of Posts and Telecommunications Science Foundation (under Grant No. NY221024).

\bibliographystyle{apsrev4-1}
\bibliography{ref}

\begin{thebibliography}{47}%
\makeatletter
\providecommand \@ifxundefined [1]{%
 \@ifx{#1\undefined}
}%
\providecommand \@ifnum [1]{%
 \ifnum #1\expandafter \@firstoftwo
 \else \expandafter \@secondoftwo
 \fi
}%
\providecommand \@ifx [1]{%
 \ifx #1\expandafter \@firstoftwo
 \else \expandafter \@secondoftwo
 \fi
}%
\providecommand \natexlab [1]{#1}%
\providecommand \enquote  [1]{``#1''}%
\providecommand \bibnamefont  [1]{#1}%
\providecommand \bibfnamefont [1]{#1}%
\providecommand \citenamefont [1]{#1}%
\providecommand \href@noop [0]{\@secondoftwo}%
\providecommand \href [0]{\begingroup \@sanitize@url \@href}%
\providecommand \@href[1]{\@@startlink{#1}\@@href}%
\providecommand \@@href[1]{\endgroup#1\@@endlink}%
\providecommand \@sanitize@url [0]{\catcode `\\12\catcode `\$12\catcode
  `\&12\catcode `\#12\catcode `\^12\catcode `\_12\catcode `\%12\relax}%
\providecommand \@@startlink[1]{}%
\providecommand \@@endlink[0]{}%
\providecommand \url  [0]{\begingroup\@sanitize@url \@url }%
\providecommand \@url [1]{\endgroup\@href {#1}{\urlprefix }}%
\providecommand \urlprefix  [0]{URL }%
\providecommand \Eprint [0]{\href }%
\providecommand \doibase [0]{http://dx.doi.org/}%
\providecommand \selectlanguage [0]{\@gobble}%
\providecommand \bibinfo  [0]{\@secondoftwo}%
\providecommand \bibfield  [0]{\@secondoftwo}%
\providecommand \translation [1]{[#1]}%
\providecommand \BibitemOpen [0]{}%
\providecommand \bibitemStop [0]{}%
\providecommand \bibitemNoStop [0]{.\EOS\space}%
\providecommand \EOS [0]{\spacefactor3000\relax}%
\providecommand \BibitemShut  [1]{\csname bibitem#1\endcsname}%
\let\auto@bib@innerbib\@empty
\bibitem [{\citenamefont {Marcal}\ \emph {et~al.}(2014)\citenamefont {Marcal},
  \citenamefont {Rosa}, \citenamefont {Safar}, \citenamefont {Freitas},
  \citenamefont {Schmidt}, \citenamefont {Guimaraes}, \citenamefont {Deneke},\
  and\ \citenamefont {Malachias}}]{ph500144s}%
  \BibitemOpen
  \bibfield  {author} {\bibinfo {author} {\bibfnamefont {L.~A.~B.}\
  \bibnamefont {Marcal}}, \bibinfo {author} {\bibfnamefont {B.~L.~T.}\
  \bibnamefont {Rosa}}, \bibinfo {author} {\bibfnamefont {G.~A.~M.}\
  \bibnamefont {Safar}}, \bibinfo {author} {\bibfnamefont {R.~O.}\ \bibnamefont
  {Freitas}}, \bibinfo {author} {\bibfnamefont {O.~G.}\ \bibnamefont
  {Schmidt}}, \bibinfo {author} {\bibfnamefont {P.~S.~S.}\ \bibnamefont
  {Guimaraes}}, \bibinfo {author} {\bibfnamefont {C.}~\bibnamefont {Deneke}}, \
  and\ \bibinfo {author} {\bibfnamefont {A.}~\bibnamefont {Malachias}},\ }\href
  {\doibase 10.1021/ph500144s} {\bibfield  {journal} {\bibinfo  {journal} {ACS
  Photonics}\ }\textbf {\bibinfo {volume} {1}},\ \bibinfo {pages} {863}
  (\bibinfo {year} {2014})}\BibitemShut {NoStop}%
\bibitem [{\citenamefont {Ren}\ and\ \citenamefont {Gao}(2014)}]{C4NR00330F}%
  \BibitemOpen
  \bibfield  {author} {\bibinfo {author} {\bibfnamefont {Z.}~\bibnamefont
  {Ren}}\ and\ \bibinfo {author} {\bibfnamefont {P.-X.}\ \bibnamefont {Gao}},\
  }\href {\doibase 10.1039/C4NR00330F} {\bibfield  {journal} {\bibinfo
  {journal} {Nanoscale}\ }\textbf {\bibinfo {volume} {6}},\ \bibinfo {pages}
  {9366} (\bibinfo {year} {2014})}\BibitemShut {NoStop}%
\bibitem [{\citenamefont {Sun}\ \emph {et~al.}(2017)\citenamefont {Sun},
  \citenamefont {Zhang}, \citenamefont {Qiu}, \citenamefont {Liu},\ and\
  \citenamefont {Xu}}]{201705630}%
  \BibitemOpen
  \bibfield  {author} {\bibinfo {author} {\bibfnamefont {Q.}~\bibnamefont
  {Sun}}, \bibinfo {author} {\bibfnamefont {R.}~\bibnamefont {Zhang}}, \bibinfo
  {author} {\bibfnamefont {J.}~\bibnamefont {Qiu}}, \bibinfo {author}
  {\bibfnamefont {R.}~\bibnamefont {Liu}}, \ and\ \bibinfo {author}
  {\bibfnamefont {W.}~\bibnamefont {Xu}},\ }\href {\doibase
  10.1002/adma.201705630} {\bibfield  {journal} {\bibinfo  {journal} {Adv.
  Mater.}\ }\textbf {\bibinfo {volume} {30}},\ \bibinfo {pages} {1705630}
  (\bibinfo {year} {2017})}\BibitemShut {NoStop}%
\bibitem [{\citenamefont {Pogosov}\ \emph {et~al.}(2022)\citenamefont
  {Pogosov}, \citenamefont {Shevyrin}, \citenamefont {Pokhabov}, \citenamefont
  {Zhdanov},\ and\ \citenamefont {Kumar}}]{Pogosov_2022}%
  \BibitemOpen
  \bibfield  {author} {\bibinfo {author} {\bibfnamefont {A.~G.}\ \bibnamefont
  {Pogosov}}, \bibinfo {author} {\bibfnamefont {A.~A.}\ \bibnamefont
  {Shevyrin}}, \bibinfo {author} {\bibfnamefont {D.~A.}\ \bibnamefont
  {Pokhabov}}, \bibinfo {author} {\bibfnamefont {E.~Y.}\ \bibnamefont
  {Zhdanov}}, \ and\ \bibinfo {author} {\bibfnamefont {S.}~\bibnamefont
  {Kumar}},\ }\href {\doibase 10.1088/1361-648x/ac6308} {\bibfield  {journal}
  {\bibinfo  {journal} {Journal of Physics: Condensed Matter}\ }\textbf
  {\bibinfo {volume} {34}},\ \bibinfo {pages} {263001} (\bibinfo {year}
  {2022})}\BibitemShut {NoStop}%
\bibitem [{\citenamefont {Schultheiss}\ \emph {et~al.}(2010)\citenamefont
  {Schultheiss}, \citenamefont {Batz}, \citenamefont {Szameit}, \citenamefont
  {Dreisow}, \citenamefont {Nolte}, \citenamefont {T\"unnermann}, \citenamefont
  {Longhi},\ and\ \citenamefont {Peschel}}]{PhysRevLett.105.143901}%
  \BibitemOpen
  \bibfield  {author} {\bibinfo {author} {\bibfnamefont {V.~H.}\ \bibnamefont
  {Schultheiss}}, \bibinfo {author} {\bibfnamefont {S.}~\bibnamefont {Batz}},
  \bibinfo {author} {\bibfnamefont {A.}~\bibnamefont {Szameit}}, \bibinfo
  {author} {\bibfnamefont {F.}~\bibnamefont {Dreisow}}, \bibinfo {author}
  {\bibfnamefont {S.}~\bibnamefont {Nolte}}, \bibinfo {author} {\bibfnamefont
  {A.}~\bibnamefont {T\"unnermann}}, \bibinfo {author} {\bibfnamefont
  {S.}~\bibnamefont {Longhi}}, \ and\ \bibinfo {author} {\bibfnamefont
  {U.}~\bibnamefont {Peschel}},\ }\href {\doibase
  10.1103/PhysRevLett.105.143901} {\bibfield  {journal} {\bibinfo  {journal}
  {Phys. Rev. Lett.}\ }\textbf {\bibinfo {volume} {105}},\ \bibinfo {pages}
  {143901} (\bibinfo {year} {2010})}\BibitemShut {NoStop}%
\bibitem [{\citenamefont {Schultheiss}\ \emph {et~al.}(2016)\citenamefont
  {Schultheiss}, \citenamefont {Batz},\ and\ \citenamefont {Peschel}}]{RN184}%
  \BibitemOpen
  \bibfield  {author} {\bibinfo {author} {\bibfnamefont {V.~H.}\ \bibnamefont
  {Schultheiss}}, \bibinfo {author} {\bibfnamefont {S.}~\bibnamefont {Batz}}, \
  and\ \bibinfo {author} {\bibfnamefont {U.}~\bibnamefont {Peschel}},\ }\href
  {\doibase 10.1038/nphoton.2015.244} {\bibfield  {journal} {\bibinfo
  {journal} {Nature Photonics}\ }\textbf {\bibinfo {volume} {10}},\ \bibinfo
  {pages} {106} (\bibinfo {year} {2016})}\BibitemShut {NoStop}%
\bibitem [{\citenamefont {Bekenstein}(2017)}]{Bekenstein2017Control}%
  \BibitemOpen
  \bibfield  {author} {\bibinfo {author} {\bibfnamefont {R.}~\bibnamefont
  {Bekenstein}},\ }\href {https://doi.org/10.1038/s41566-017-0008-0} {\bibfield
   {journal} {\bibinfo  {journal} {Nature Photon.}\ }\textbf {\bibinfo {volume}
  {11}} (\bibinfo {year} {2017})}\BibitemShut {NoStop}%
\bibitem [{\citenamefont {Zhu}\ \emph {et~al.}(2018)\citenamefont {Zhu},
  \citenamefont {Liu}, \citenamefont {Liang}, \citenamefont {Chen},\ and\
  \citenamefont {Li}}]{PhysRevLett.121.234301}%
  \BibitemOpen
  \bibfield  {author} {\bibinfo {author} {\bibfnamefont {J.}~\bibnamefont
  {Zhu}}, \bibinfo {author} {\bibfnamefont {Y.}~\bibnamefont {Liu}}, \bibinfo
  {author} {\bibfnamefont {Z.}~\bibnamefont {Liang}}, \bibinfo {author}
  {\bibfnamefont {T.}~\bibnamefont {Chen}}, \ and\ \bibinfo {author}
  {\bibfnamefont {J.}~\bibnamefont {Li}},\ }\href {\doibase
  10.1103/PhysRevLett.121.234301} {\bibfield  {journal} {\bibinfo  {journal}
  {Phys. Rev. Lett.}\ }\textbf {\bibinfo {volume} {121}},\ \bibinfo {pages}
  {234301} (\bibinfo {year} {2018})}\BibitemShut {NoStop}%
\bibitem [{\citenamefont {Mcmahon}\ and\ \citenamefont
  {Gallop}(2005)}]{2005Membrane}%
  \BibitemOpen
  \bibfield  {author} {\bibinfo {author} {\bibfnamefont {H.~T.}\ \bibnamefont
  {Mcmahon}}\ and\ \bibinfo {author} {\bibfnamefont {J.~L.}\ \bibnamefont
  {Gallop}},\ }\href {https://doi.org/10.1038/nature04396} {\bibfield
  {journal} {\bibinfo  {journal} {Nature}\ }\textbf {\bibinfo {volume} {438}},\
  \bibinfo {pages} {590} (\bibinfo {year} {2005})}\BibitemShut {NoStop}%
\bibitem [{\citenamefont {Pieuchot}\ \emph {et~al.}(2018)\citenamefont
  {Pieuchot}, \citenamefont {Marteau}, \citenamefont {Guignandon},
  \citenamefont {Dos~Santos}, \citenamefont {Brigaud}, \citenamefont {Chauvy},
  \citenamefont {Cloatre}, \citenamefont {Ponche}, \citenamefont {Petithory},
  \citenamefont {Rougerie}, \citenamefont {Vassaux}, \citenamefont {Milan},
  \citenamefont {Tusamda~Wakhloo}, \citenamefont {Spangenberg}, \citenamefont
  {Bigerelle},\ and\ \citenamefont {Anselme}}]{RN1}%
  \BibitemOpen
  \bibfield  {author} {\bibinfo {author} {\bibfnamefont {L.}~\bibnamefont
  {Pieuchot}}, \bibinfo {author} {\bibfnamefont {J.}~\bibnamefont {Marteau}},
  \bibinfo {author} {\bibfnamefont {A.}~\bibnamefont {Guignandon}}, \bibinfo
  {author} {\bibfnamefont {T.}~\bibnamefont {Dos~Santos}}, \bibinfo {author}
  {\bibfnamefont {I.}~\bibnamefont {Brigaud}}, \bibinfo {author} {\bibfnamefont
  {P.-F.}\ \bibnamefont {Chauvy}}, \bibinfo {author} {\bibfnamefont
  {T.}~\bibnamefont {Cloatre}}, \bibinfo {author} {\bibfnamefont
  {A.}~\bibnamefont {Ponche}}, \bibinfo {author} {\bibfnamefont
  {T.}~\bibnamefont {Petithory}}, \bibinfo {author} {\bibfnamefont
  {P.}~\bibnamefont {Rougerie}}, \bibinfo {author} {\bibfnamefont
  {M.}~\bibnamefont {Vassaux}}, \bibinfo {author} {\bibfnamefont {J.-L.}\
  \bibnamefont {Milan}}, \bibinfo {author} {\bibfnamefont {N.}~\bibnamefont
  {Tusamda~Wakhloo}}, \bibinfo {author} {\bibfnamefont {A.}~\bibnamefont
  {Spangenberg}}, \bibinfo {author} {\bibfnamefont {M.}~\bibnamefont
  {Bigerelle}}, \ and\ \bibinfo {author} {\bibfnamefont {K.}~\bibnamefont
  {Anselme}},\ }\href {\doibase 10.1038/s41467-018-06494-6} {\bibfield
  {journal} {\bibinfo  {journal} {Nature Communications}\ }\textbf {\bibinfo
  {volume} {9}},\ \bibinfo {pages} {3995} (\bibinfo {year} {2018})}\BibitemShut
  {NoStop}%
\bibitem [{\citenamefont {Nishide}\ and\ \citenamefont
  {Ishihara}(2022)}]{PhysRevLett.128.224101}%
  \BibitemOpen
  \bibfield  {author} {\bibinfo {author} {\bibfnamefont {R.}~\bibnamefont
  {Nishide}}\ and\ \bibinfo {author} {\bibfnamefont {S.}~\bibnamefont
  {Ishihara}},\ }\href {\doibase 10.1103/PhysRevLett.128.224101} {\bibfield
  {journal} {\bibinfo  {journal} {Phys. Rev. Lett.}\ }\textbf {\bibinfo
  {volume} {128}},\ \bibinfo {pages} {224101} (\bibinfo {year}
  {2022})}\BibitemShut {NoStop}%
\bibitem [{\citenamefont {Maroudas-Sacks}\ \emph {et~al.}(2021)\citenamefont
  {Maroudas-Sacks}, \citenamefont {Garion}, \citenamefont {Shani-Zerbib},
  \citenamefont {Livshits}, \citenamefont {Braun},\ and\ \citenamefont
  {Keren}}]{RN2}%
  \BibitemOpen
  \bibfield  {author} {\bibinfo {author} {\bibfnamefont {Y.}~\bibnamefont
  {Maroudas-Sacks}}, \bibinfo {author} {\bibfnamefont {L.}~\bibnamefont
  {Garion}}, \bibinfo {author} {\bibfnamefont {L.}~\bibnamefont
  {Shani-Zerbib}}, \bibinfo {author} {\bibfnamefont {A.}~\bibnamefont
  {Livshits}}, \bibinfo {author} {\bibfnamefont {E.}~\bibnamefont {Braun}}, \
  and\ \bibinfo {author} {\bibfnamefont {K.}~\bibnamefont {Keren}},\ }\href
  {\doibase 10.1038/s41567-020-01083-1} {\bibfield  {journal} {\bibinfo
  {journal} {Nature Physics}\ }\textbf {\bibinfo {volume} {17}},\ \bibinfo
  {pages} {251} (\bibinfo {year} {2021})}\BibitemShut {NoStop}%
\bibitem [{\citenamefont {Horibe}\ \emph {et~al.}(2019)\citenamefont {Horibe},
  \citenamefont {Hironaka}, \citenamefont {Matsushita},\ and\ \citenamefont
  {Fujimoto}}]{1.5108838}%
  \BibitemOpen
  \bibfield  {author} {\bibinfo {author} {\bibfnamefont {K.}~\bibnamefont
  {Horibe}}, \bibinfo {author} {\bibfnamefont {K.-i.}\ \bibnamefont
  {Hironaka}}, \bibinfo {author} {\bibfnamefont {K.}~\bibnamefont
  {Matsushita}}, \ and\ \bibinfo {author} {\bibfnamefont {K.}~\bibnamefont
  {Fujimoto}},\ }\href {\doibase 10.1063/1.5108838} {\bibfield  {journal}
  {\bibinfo  {journal} {Chaos: An Interdisciplinary Journal of Nonlinear
  Science}\ }\textbf {\bibinfo {volume} {29}},\ \bibinfo {pages} {093120}
  (\bibinfo {year} {2019})}\BibitemShut {NoStop}%
\bibitem [{\citenamefont {Saito}\ and\ \citenamefont
  {Sawai}(2021)}]{SAITO2021103087}%
  \BibitemOpen
  \bibfield  {author} {\bibinfo {author} {\bibfnamefont {N.}~\bibnamefont
  {Saito}}\ and\ \bibinfo {author} {\bibfnamefont {S.}~\bibnamefont {Sawai}},\
  }\href {\doibase https://doi.org/10.1016/j.isci.2021.103087} {\bibfield
  {journal} {\bibinfo  {journal} {iScience}\ }\textbf {\bibinfo {volume}
  {24}},\ \bibinfo {pages} {103087} (\bibinfo {year} {2021})}\BibitemShut
  {NoStop}%
\bibitem [{\citenamefont {Ortix}\ \emph {et~al.}(2011)\citenamefont {Ortix},
  \citenamefont {Kiravittaya}, \citenamefont {Schmidt},\ and\ \citenamefont
  {van~den Brink}}]{PhysRevB.84.045438}%
  \BibitemOpen
  \bibfield  {author} {\bibinfo {author} {\bibfnamefont {C.}~\bibnamefont
  {Ortix}}, \bibinfo {author} {\bibfnamefont {S.}~\bibnamefont {Kiravittaya}},
  \bibinfo {author} {\bibfnamefont {O.~G.}\ \bibnamefont {Schmidt}}, \ and\
  \bibinfo {author} {\bibfnamefont {J.}~\bibnamefont {van~den Brink}},\ }\href
  {\doibase 10.1103/PhysRevB.84.045438} {\bibfield  {journal} {\bibinfo
  {journal} {Phys. Rev. B}\ }\textbf {\bibinfo {volume} {84}},\ \bibinfo
  {pages} {045438} (\bibinfo {year} {2011})}\BibitemShut {NoStop}%
\bibitem [{\citenamefont {Fomin}\ \emph {et~al.}(2012)\citenamefont {Fomin},
  \citenamefont {Kiravittaya},\ and\ \citenamefont
  {Schmidt}}]{PhysRevB.86.195421}%
  \BibitemOpen
  \bibfield  {author} {\bibinfo {author} {\bibfnamefont {V.~M.}\ \bibnamefont
  {Fomin}}, \bibinfo {author} {\bibfnamefont {S.}~\bibnamefont {Kiravittaya}},
  \ and\ \bibinfo {author} {\bibfnamefont {O.~G.}\ \bibnamefont {Schmidt}},\
  }\href {\doibase 10.1103/PhysRevB.86.195421} {\bibfield  {journal} {\bibinfo
  {journal} {Phys. Rev. B}\ }\textbf {\bibinfo {volume} {86}},\ \bibinfo
  {pages} {195421} (\bibinfo {year} {2012})}\BibitemShut {NoStop}%
\bibitem [{\citenamefont {Chang}\ and\ \citenamefont
  {Ortix}(2017)}]{Chang_2017}%
  \BibitemOpen
  \bibfield  {author} {\bibinfo {author} {\bibfnamefont {C.-H.}\ \bibnamefont
  {Chang}}\ and\ \bibinfo {author} {\bibfnamefont {C.}~\bibnamefont {Ortix}},\
  }\href {\doibase 10.1088/2399-1984/aa9c7a} {\bibfield  {journal} {\bibinfo
  {journal} {Nano Futures}\ }\textbf {\bibinfo {volume} {1}},\ \bibinfo {pages}
  {035004} (\bibinfo {year} {2017})}\BibitemShut {NoStop}%
\bibitem [{\citenamefont {Streubel}\ \emph {et~al.}(2016)\citenamefont
  {Streubel}, \citenamefont {Kronast}, \citenamefont {Reiche}, \citenamefont
  {Mühl}, \citenamefont {Wolter}, \citenamefont {Schmidt},\ and\ \citenamefont
  {Makarov}}]{1.4941045}%
  \BibitemOpen
  \bibfield  {author} {\bibinfo {author} {\bibfnamefont {R.}~\bibnamefont
  {Streubel}}, \bibinfo {author} {\bibfnamefont {F.}~\bibnamefont {Kronast}},
  \bibinfo {author} {\bibfnamefont {C.~F.}\ \bibnamefont {Reiche}}, \bibinfo
  {author} {\bibfnamefont {T.}~\bibnamefont {Mühl}}, \bibinfo {author}
  {\bibfnamefont {A.~U.~B.}\ \bibnamefont {Wolter}}, \bibinfo {author}
  {\bibfnamefont {O.~G.}\ \bibnamefont {Schmidt}}, \ and\ \bibinfo {author}
  {\bibfnamefont {D.}~\bibnamefont {Makarov}},\ }\href {\doibase
  10.1063/1.4941045} {\bibfield  {journal} {\bibinfo  {journal} {Applied
  Physics Letters}\ }\textbf {\bibinfo {volume} {108}},\ \bibinfo {pages}
  {042407} (\bibinfo {year} {2016})}\BibitemShut {NoStop}%
\bibitem [{\citenamefont {Makarov}\ \emph {et~al.}(2022)\citenamefont
  {Makarov}, \citenamefont {Volkov}, \citenamefont {Kákay}, \citenamefont
  {Pylypovskyi}, \citenamefont {Budinská},\ and\ \citenamefont
  {Dobrovolskiy}}]{adma.202101758}%
  \BibitemOpen
  \bibfield  {author} {\bibinfo {author} {\bibfnamefont {D.}~\bibnamefont
  {Makarov}}, \bibinfo {author} {\bibfnamefont {O.~M.}\ \bibnamefont {Volkov}},
  \bibinfo {author} {\bibfnamefont {A.}~\bibnamefont {Kákay}}, \bibinfo
  {author} {\bibfnamefont {O.~V.}\ \bibnamefont {Pylypovskyi}}, \bibinfo
  {author} {\bibfnamefont {B.}~\bibnamefont {Budinská}}, \ and\ \bibinfo
  {author} {\bibfnamefont {O.~V.}\ \bibnamefont {Dobrovolskiy}},\ }\href
  {\doibase https://doi.org/10.1002/adma.202101758} {\bibfield  {journal}
  {\bibinfo  {journal} {Advanced Materials}\ }\textbf {\bibinfo {volume}
  {34}},\ \bibinfo {pages} {2101758} (\bibinfo {year} {2022})}\BibitemShut
  {NoStop}%
\bibitem [{\citenamefont {Tempere}\ \emph {et~al.}(2009)\citenamefont
  {Tempere}, \citenamefont {Gladilin}, \citenamefont {Silvera}, \citenamefont
  {Devreese},\ and\ \citenamefont {Moshchalkov}}]{PhysRevB.79.134516}%
  \BibitemOpen
  \bibfield  {author} {\bibinfo {author} {\bibfnamefont {J.}~\bibnamefont
  {Tempere}}, \bibinfo {author} {\bibfnamefont {V.~N.}\ \bibnamefont
  {Gladilin}}, \bibinfo {author} {\bibfnamefont {I.~F.}\ \bibnamefont
  {Silvera}}, \bibinfo {author} {\bibfnamefont {J.~T.}\ \bibnamefont
  {Devreese}}, \ and\ \bibinfo {author} {\bibfnamefont {V.~V.}\ \bibnamefont
  {Moshchalkov}},\ }\href {\doibase 10.1103/PhysRevB.79.134516} {\bibfield
  {journal} {\bibinfo  {journal} {Phys. Rev. B}\ }\textbf {\bibinfo {volume}
  {79}},\ \bibinfo {pages} {134516} (\bibinfo {year} {2009})}\BibitemShut
  {NoStop}%
\bibitem [{\citenamefont {Ying}\ \emph {et~al.}(2016)\citenamefont {Ying},
  \citenamefont {Gentile}, \citenamefont {Ortix},\ and\ \citenamefont
  {Cuoco}}]{PhysRevB.94.081406}%
  \BibitemOpen
  \bibfield  {author} {\bibinfo {author} {\bibfnamefont {Z.-J.}\ \bibnamefont
  {Ying}}, \bibinfo {author} {\bibfnamefont {P.}~\bibnamefont {Gentile}},
  \bibinfo {author} {\bibfnamefont {C.}~\bibnamefont {Ortix}}, \ and\ \bibinfo
  {author} {\bibfnamefont {M.}~\bibnamefont {Cuoco}},\ }\href {\doibase
  10.1103/PhysRevB.94.081406} {\bibfield  {journal} {\bibinfo  {journal} {Phys.
  Rev. B}\ }\textbf {\bibinfo {volume} {94}},\ \bibinfo {pages} {081406}
  (\bibinfo {year} {2016})}\BibitemShut {NoStop}%
\bibitem [{\citenamefont {Law}\ \emph {et~al.}(2020)\citenamefont {Law},
  \citenamefont {Dean}, \citenamefont {Miller},\ and\ \citenamefont
  {Kusumaatmaja}}]{D0SM00652A}%
  \BibitemOpen
  \bibfield  {author} {\bibinfo {author} {\bibfnamefont {J.~O.}\ \bibnamefont
  {Law}}, \bibinfo {author} {\bibfnamefont {J.~M.}\ \bibnamefont {Dean}},
  \bibinfo {author} {\bibfnamefont {M.~A.}\ \bibnamefont {Miller}}, \ and\
  \bibinfo {author} {\bibfnamefont {H.}~\bibnamefont {Kusumaatmaja}},\ }\href
  {\doibase 10.1039/D0SM00652A} {\bibfield  {journal} {\bibinfo  {journal}
  {Soft Matter}\ }\textbf {\bibinfo {volume} {16}},\ \bibinfo {pages} {8069}
  (\bibinfo {year} {2020})}\BibitemShut {NoStop}%
\bibitem [{\citenamefont {Jensen}\ and\ \citenamefont
  {Koppe}(1971)}]{JENSEN1971586}%
  \BibitemOpen
  \bibfield  {author} {\bibinfo {author} {\bibfnamefont {H.}~\bibnamefont
  {Jensen}}\ and\ \bibinfo {author} {\bibfnamefont {H.}~\bibnamefont {Koppe}},\
  }\href {\doibase https://doi.org/10.1016/0003-4916(71)90031-5} {\bibfield
  {journal} {\bibinfo  {journal} {Ann. Phys.}\ }\textbf {\bibinfo {volume}
  {63}},\ \bibinfo {pages} {586 } (\bibinfo {year} {1971})}\BibitemShut
  {NoStop}%
\bibitem [{\citenamefont {da~Costa}(1981)}]{PhysRevA.23.1982}%
  \BibitemOpen
  \bibfield  {author} {\bibinfo {author} {\bibfnamefont {R.~C.~T.}\
  \bibnamefont {da~Costa}},\ }\href {\doibase 10.1103/PhysRevA.23.1982}
  {\bibfield  {journal} {\bibinfo  {journal} {Phys. Rev. A}\ }\textbf {\bibinfo
  {volume} {23}},\ \bibinfo {pages} {1982} (\bibinfo {year}
  {1981})}\BibitemShut {NoStop}%
\bibitem [{\citenamefont {Szameit}\ \emph {et~al.}(2010)\citenamefont
  {Szameit}, \citenamefont {Dreisow}, \citenamefont {Heinrich}, \citenamefont
  {Keil}, \citenamefont {Nolte}, \citenamefont {T\"unnermann},\ and\
  \citenamefont {Longhi}}]{PhysRevLett.104.150403}%
  \BibitemOpen
  \bibfield  {author} {\bibinfo {author} {\bibfnamefont {A.}~\bibnamefont
  {Szameit}}, \bibinfo {author} {\bibfnamefont {F.}~\bibnamefont {Dreisow}},
  \bibinfo {author} {\bibfnamefont {M.}~\bibnamefont {Heinrich}}, \bibinfo
  {author} {\bibfnamefont {R.}~\bibnamefont {Keil}}, \bibinfo {author}
  {\bibfnamefont {S.}~\bibnamefont {Nolte}}, \bibinfo {author} {\bibfnamefont
  {A.}~\bibnamefont {T\"unnermann}}, \ and\ \bibinfo {author} {\bibfnamefont
  {S.}~\bibnamefont {Longhi}},\ }\href {\doibase
  10.1103/PhysRevLett.104.150403} {\bibfield  {journal} {\bibinfo  {journal}
  {Phys. Rev. Lett.}\ }\textbf {\bibinfo {volume} {104}},\ \bibinfo {pages}
  {150403} (\bibinfo {year} {2010})}\BibitemShut {NoStop}%
\bibitem [{\citenamefont {Ferrari}\ and\ \citenamefont
  {Cuoghi}(2008)}]{PhysRevLett.100.230403}%
  \BibitemOpen
  \bibfield  {author} {\bibinfo {author} {\bibfnamefont {G.}~\bibnamefont
  {Ferrari}}\ and\ \bibinfo {author} {\bibfnamefont {G.}~\bibnamefont
  {Cuoghi}},\ }\href {\doibase 10.1103/PhysRevLett.100.230403} {\bibfield
  {journal} {\bibinfo  {journal} {Phys. Rev. Lett.}\ }\textbf {\bibinfo
  {volume} {100}},\ \bibinfo {pages} {230403} (\bibinfo {year}
  {2008})}\BibitemShut {NoStop}%
\bibitem [{\citenamefont {Brandt}\ and\ \citenamefont
  {S{\'{a}}nchez-Monroy}(2015)}]{Brandt_2015}%
  \BibitemOpen
  \bibfield  {author} {\bibinfo {author} {\bibfnamefont {F.~T.}\ \bibnamefont
  {Brandt}}\ and\ \bibinfo {author} {\bibfnamefont {J.~A.}\ \bibnamefont
  {S{\'{a}}nchez-Monroy}},\ }\href {\doibase 10.1209/0295-5075/111/67004}
  {\bibfield  {journal} {\bibinfo  {journal} {EPL}\ }\textbf {\bibinfo {volume}
  {111}},\ \bibinfo {pages} {67004} (\bibinfo {year} {2015})}\BibitemShut
  {NoStop}%
\bibitem [{\citenamefont {Wang}\ and\ \citenamefont {Zong}(2016)}]{Wang2016a}%
  \BibitemOpen
  \bibfield  {author} {\bibinfo {author} {\bibfnamefont {Y.-L.}\ \bibnamefont
  {Wang}}\ and\ \bibinfo {author} {\bibfnamefont {H.-S.}\ \bibnamefont
  {Zong}},\ }\href {\doibase https://doi.org/10.1016/j.aop.2015.10.019}
  {\bibfield  {journal} {\bibinfo  {journal} {Ann. Phys.}\ }\textbf {\bibinfo
  {volume} {364}},\ \bibinfo {pages} {68 } (\bibinfo {year}
  {2016})}\BibitemShut {NoStop}%
\bibitem [{\citenamefont {Ouyang}\ \emph {et~al.}(1999)\citenamefont {Ouyang},
  \citenamefont {Mohta},\ and\ \citenamefont {Jaffe}}]{OUYANG1999297}%
  \BibitemOpen
  \bibfield  {author} {\bibinfo {author} {\bibfnamefont {P.}~\bibnamefont
  {Ouyang}}, \bibinfo {author} {\bibfnamefont {V.}~\bibnamefont {Mohta}}, \
  and\ \bibinfo {author} {\bibfnamefont {R.}~\bibnamefont {Jaffe}},\ }\href
  {\doibase https://doi.org/10.1006/aphy.1999.5935} {\bibfield  {journal}
  {\bibinfo  {journal} {Ann. Phys.}\ }\textbf {\bibinfo {volume} {275}},\
  \bibinfo {pages} {297 } (\bibinfo {year} {1999})}\BibitemShut {NoStop}%
\bibitem [{\citenamefont {Burgess}\ and\ \citenamefont
  {Jensen}(1993)}]{PhysRevA.48.1861}%
  \BibitemOpen
  \bibfield  {author} {\bibinfo {author} {\bibfnamefont {M.}~\bibnamefont
  {Burgess}}\ and\ \bibinfo {author} {\bibfnamefont {B.}~\bibnamefont
  {Jensen}},\ }\href {\doibase 10.1103/PhysRevA.48.1861} {\bibfield  {journal}
  {\bibinfo  {journal} {Phys. Rev. A}\ }\textbf {\bibinfo {volume} {48}},\
  \bibinfo {pages} {1861} (\bibinfo {year} {1993})}\BibitemShut {NoStop}%
\bibitem [{\citenamefont {Brandt}\ and\ \citenamefont
  {Sánchez-Monroy}(2016)}]{BRANDT20163036}%
  \BibitemOpen
  \bibfield  {author} {\bibinfo {author} {\bibfnamefont {F.}~\bibnamefont
  {Brandt}}\ and\ \bibinfo {author} {\bibfnamefont {J.}~\bibnamefont
  {Sánchez-Monroy}},\ }\href {\doibase
  https://doi.org/10.1016/j.physleta.2016.07.010} {\bibfield  {journal}
  {\bibinfo  {journal} {Phys. Lett. A}\ }\textbf {\bibinfo {volume} {380}},\
  \bibinfo {pages} {3036 } (\bibinfo {year} {2016})}\BibitemShut {NoStop}%
\bibitem [{\citenamefont {Ortix}(2015)}]{PhysRevB.91.245412}%
  \BibitemOpen
  \bibfield  {author} {\bibinfo {author} {\bibfnamefont {C.}~\bibnamefont
  {Ortix}},\ }\href {\doibase 10.1103/PhysRevB.91.245412} {\bibfield  {journal}
  {\bibinfo  {journal} {Phys. Rev. B}\ }\textbf {\bibinfo {volume} {91}},\
  \bibinfo {pages} {245412} (\bibinfo {year} {2015})}\BibitemShut {NoStop}%
\bibitem [{\citenamefont {Entin}\ and\ \citenamefont
  {Magarill}(2001)}]{PhysRevB.64.085330}%
  \BibitemOpen
  \bibfield  {author} {\bibinfo {author} {\bibfnamefont {M.~V.}\ \bibnamefont
  {Entin}}\ and\ \bibinfo {author} {\bibfnamefont {L.~I.}\ \bibnamefont
  {Magarill}},\ }\href {\doibase 10.1103/PhysRevB.64.085330} {\bibfield
  {journal} {\bibinfo  {journal} {Phys. Rev. B}\ }\textbf {\bibinfo {volume}
  {64}},\ \bibinfo {pages} {085330} (\bibinfo {year} {2001})}\BibitemShut
  {NoStop}%
\bibitem [{\citenamefont {Chang}\ \emph {et~al.}(2013)\citenamefont {Chang},
  \citenamefont {Wu},\ and\ \citenamefont {Chang}}]{PhysRevB.87.174413}%
  \BibitemOpen
  \bibfield  {author} {\bibinfo {author} {\bibfnamefont {J.-Y.}\ \bibnamefont
  {Chang}}, \bibinfo {author} {\bibfnamefont {J.-S.}\ \bibnamefont {Wu}}, \
  and\ \bibinfo {author} {\bibfnamefont {C.-R.}\ \bibnamefont {Chang}},\ }\href
  {\doibase 10.1103/PhysRevB.87.174413} {\bibfield  {journal} {\bibinfo
  {journal} {Phys. Rev. B}\ }\textbf {\bibinfo {volume} {87}},\ \bibinfo
  {pages} {174413} (\bibinfo {year} {2013})}\BibitemShut {NoStop}%
\bibitem [{\citenamefont {Wang}\ \emph {et~al.}(2014)\citenamefont {Wang},
  \citenamefont {Du}, \citenamefont {Xu}, \citenamefont {Liu},\ and\
  \citenamefont {Zong}}]{PhysRevA.90.042117}%
  \BibitemOpen
  \bibfield  {author} {\bibinfo {author} {\bibfnamefont {Y.-L.}\ \bibnamefont
  {Wang}}, \bibinfo {author} {\bibfnamefont {L.}~\bibnamefont {Du}}, \bibinfo
  {author} {\bibfnamefont {C.-T.}\ \bibnamefont {Xu}}, \bibinfo {author}
  {\bibfnamefont {X.-J.}\ \bibnamefont {Liu}}, \ and\ \bibinfo {author}
  {\bibfnamefont {H.-S.}\ \bibnamefont {Zong}},\ }\href {\doibase
  10.1103/PhysRevA.90.042117} {\bibfield  {journal} {\bibinfo  {journal} {Phys.
  Rev. A}\ }\textbf {\bibinfo {volume} {90}},\ \bibinfo {pages} {042117}
  (\bibinfo {year} {2014})}\BibitemShut {NoStop}%
\bibitem [{\citenamefont {Wang}\ \emph {et~al.}(2017)\citenamefont {Wang},
  \citenamefont {Jiang},\ and\ \citenamefont {Zong}}]{Wang2017}%
  \BibitemOpen
  \bibfield  {author} {\bibinfo {author} {\bibfnamefont {Y.-L.}\ \bibnamefont
  {Wang}}, \bibinfo {author} {\bibfnamefont {H.}~\bibnamefont {Jiang}}, \ and\
  \bibinfo {author} {\bibfnamefont {H.-S.}\ \bibnamefont {Zong}},\ }\href
  {\doibase 10.1103/PhysRevA.96.022116} {\bibfield  {journal} {\bibinfo
  {journal} {Phys. Rev. A}\ }\textbf {\bibinfo {volume} {96}},\ \bibinfo
  {pages} {022116} (\bibinfo {year} {2017})}\BibitemShut {NoStop}%
\bibitem [{\citenamefont {Liang}\ \emph {et~al.}(2018)\citenamefont {Liang},
  \citenamefont {Wang}, \citenamefont {Lai}, \citenamefont {Liu}, \citenamefont
  {Zong},\ and\ \citenamefont {Zhu}}]{PhysRevA.98.062112}%
  \BibitemOpen
  \bibfield  {author} {\bibinfo {author} {\bibfnamefont {G.-H.}\ \bibnamefont
  {Liang}}, \bibinfo {author} {\bibfnamefont {Y.-L.}\ \bibnamefont {Wang}},
  \bibinfo {author} {\bibfnamefont {M.-Y.}\ \bibnamefont {Lai}}, \bibinfo
  {author} {\bibfnamefont {H.}~\bibnamefont {Liu}}, \bibinfo {author}
  {\bibfnamefont {H.-S.}\ \bibnamefont {Zong}}, \ and\ \bibinfo {author}
  {\bibfnamefont {S.-N.}\ \bibnamefont {Zhu}},\ }\href {\doibase
  10.1103/PhysRevA.98.062112} {\bibfield  {journal} {\bibinfo  {journal} {Phys.
  Rev. A}\ }\textbf {\bibinfo {volume} {98}},\ \bibinfo {pages} {062112}
  (\bibinfo {year} {2018})}\BibitemShut {NoStop}%
\bibitem [{\citenamefont {Batz}\ and\ \citenamefont
  {Peschel}(2008)}]{PhysRevA.78.043821}%
  \BibitemOpen
  \bibfield  {author} {\bibinfo {author} {\bibfnamefont {S.}~\bibnamefont
  {Batz}}\ and\ \bibinfo {author} {\bibfnamefont {U.}~\bibnamefont {Peschel}},\
  }\href {\doibase 10.1103/PhysRevA.78.043821} {\bibfield  {journal} {\bibinfo
  {journal} {Phys. Rev. A}\ }\textbf {\bibinfo {volume} {78}},\ \bibinfo
  {pages} {043821} (\bibinfo {year} {2008})}\BibitemShut {NoStop}%
\bibitem [{\citenamefont {Lai}\ \emph {et~al.}(2018)\citenamefont {Lai},
  \citenamefont {Wang}, \citenamefont {Liang}, \citenamefont {Wang},\ and\
  \citenamefont {Zong}}]{PhysRevA.97.033843}%
  \BibitemOpen
  \bibfield  {author} {\bibinfo {author} {\bibfnamefont {M.-Y.}\ \bibnamefont
  {Lai}}, \bibinfo {author} {\bibfnamefont {Y.-L.}\ \bibnamefont {Wang}},
  \bibinfo {author} {\bibfnamefont {G.-H.}\ \bibnamefont {Liang}}, \bibinfo
  {author} {\bibfnamefont {F.}~\bibnamefont {Wang}}, \ and\ \bibinfo {author}
  {\bibfnamefont {H.-S.}\ \bibnamefont {Zong}},\ }\href {\doibase
  10.1103/PhysRevA.97.033843} {\bibfield  {journal} {\bibinfo  {journal} {Phys.
  Rev. A}\ }\textbf {\bibinfo {volume} {97}},\ \bibinfo {pages} {033843}
  (\bibinfo {year} {2018})}\BibitemShut {NoStop}%
\bibitem [{\citenamefont {Lai}\ \emph {et~al.}(2019)\citenamefont {Lai},
  \citenamefont {Wang}, \citenamefont {Liang},\ and\ \citenamefont
  {Zong}}]{PhysRevA.100.033825}%
  \BibitemOpen
  \bibfield  {author} {\bibinfo {author} {\bibfnamefont {M.-Y.}\ \bibnamefont
  {Lai}}, \bibinfo {author} {\bibfnamefont {Y.-L.}\ \bibnamefont {Wang}},
  \bibinfo {author} {\bibfnamefont {G.-H.}\ \bibnamefont {Liang}}, \ and\
  \bibinfo {author} {\bibfnamefont {H.-S.}\ \bibnamefont {Zong}},\ }\href
  {\doibase 10.1103/PhysRevA.100.033825} {\bibfield  {journal} {\bibinfo
  {journal} {Phys. Rev. A}\ }\textbf {\bibinfo {volume} {100}},\ \bibinfo
  {pages} {033825} (\bibinfo {year} {2019})}\BibitemShut {NoStop}%
\bibitem [{\citenamefont {Maraner}\ and\ \citenamefont
  {Destri}(1993)}]{S0217732393000891}%
  \BibitemOpen
  \bibfield  {author} {\bibinfo {author} {\bibfnamefont {P.}~\bibnamefont
  {Maraner}}\ and\ \bibinfo {author} {\bibfnamefont {C.}~\bibnamefont
  {Destri}},\ }\href {\doibase 10.1142/S0217732393000891} {\bibfield  {journal}
  {\bibinfo  {journal} {Mod. Phys. Lett. A}\ }\textbf {\bibinfo {volume}
  {08}},\ \bibinfo {pages} {861} (\bibinfo {year} {1993})}\BibitemShut
  {NoStop}%
\bibitem [{\citenamefont {Maraner}(1995)}]{Maraner_1995}%
  \BibitemOpen
  \bibfield  {author} {\bibinfo {author} {\bibfnamefont {P.}~\bibnamefont
  {Maraner}},\ }\href {\doibase 10.1088/0305-4470/28/10/021} {\bibfield
  {journal} {\bibinfo  {journal} {J. Phys. A: Math. Gen.}\ }\textbf {\bibinfo
  {volume} {28}},\ \bibinfo {pages} {2939} (\bibinfo {year}
  {1995})}\BibitemShut {NoStop}%
\bibitem [{\citenamefont {Maraner}(1996)}]{MARANER1996325}%
  \BibitemOpen
  \bibfield  {author} {\bibinfo {author} {\bibfnamefont {P.}~\bibnamefont
  {Maraner}},\ }\href {\doibase https://doi.org/10.1006/aphy.1996.0029}
  {\bibfield  {journal} {\bibinfo  {journal} {Ann. Phys.}\ }\textbf {\bibinfo
  {volume} {246}},\ \bibinfo {pages} {325 } (\bibinfo {year}
  {1996})}\BibitemShut {NoStop}%
\bibitem [{\citenamefont {Fujii}\ \emph {et~al.}(1997)\citenamefont {Fujii},
  \citenamefont {Ogawa}, \citenamefont {Uchiyama},\ and\ \citenamefont
  {Chepilko}}]{S0217751X97002814}%
  \BibitemOpen
  \bibfield  {author} {\bibinfo {author} {\bibfnamefont {K.}~\bibnamefont
  {Fujii}}, \bibinfo {author} {\bibfnamefont {N.}~\bibnamefont {Ogawa}},
  \bibinfo {author} {\bibfnamefont {S.}~\bibnamefont {Uchiyama}}, \ and\
  \bibinfo {author} {\bibfnamefont {N.~M.}\ \bibnamefont {Chepilko}},\ }\href
  {\doibase 10.1142/S0217751X97002814} {\bibfield  {journal} {\bibinfo
  {journal} {Int. J. Mod. Phys. A}\ }\textbf {\bibinfo {volume} {12}},\
  \bibinfo {pages} {5235} (\bibinfo {year} {1997})}\BibitemShut {NoStop}%
\bibitem [{\citenamefont {Schuster}\ and\ \citenamefont
  {Jaffe}(2003)}]{SCHUSTER2003132}%
  \BibitemOpen
  \bibfield  {author} {\bibinfo {author} {\bibfnamefont {P.}~\bibnamefont
  {Schuster}}\ and\ \bibinfo {author} {\bibfnamefont {R.}~\bibnamefont
  {Jaffe}},\ }\href {\doibase https://doi.org/10.1016/S0003-4916(03)00080-0}
  {\bibfield  {journal} {\bibinfo  {journal} {Ann. Phys.}\ }\textbf {\bibinfo
  {volume} {307}},\ \bibinfo {pages} {132 } (\bibinfo {year}
  {2003})}\BibitemShut {NoStop}%
\bibitem [{\citenamefont {Sheng}\ \emph {et~al.}(2013)\citenamefont {Sheng},
  \citenamefont {Liu}, \citenamefont {Wang}, \citenamefont {Zhu},\ and\
  \citenamefont {Genov}}]{RN3}%
  \BibitemOpen
  \bibfield  {author} {\bibinfo {author} {\bibfnamefont {C.}~\bibnamefont
  {Sheng}}, \bibinfo {author} {\bibfnamefont {H.}~\bibnamefont {Liu}}, \bibinfo
  {author} {\bibfnamefont {Y.}~\bibnamefont {Wang}}, \bibinfo {author}
  {\bibfnamefont {S.~N.}\ \bibnamefont {Zhu}}, \ and\ \bibinfo {author}
  {\bibfnamefont {D.~A.}\ \bibnamefont {Genov}},\ }\href {\doibase
  10.1038/nphoton.2013.247} {\bibfield  {journal} {\bibinfo  {journal} {Nature
  Photonics}\ }\textbf {\bibinfo {volume} {7}},\ \bibinfo {pages} {902}
  (\bibinfo {year} {2013})}\BibitemShut {NoStop}%
\bibitem [{\citenamefont {Landauer}(1987)}]{Landauer1987}%
  \BibitemOpen
  \bibfield  {author} {\bibinfo {author} {\bibfnamefont {R.}~\bibnamefont
  {Landauer}},\ }\href {\doibase 10.1007/BF01304229} {\bibfield  {journal}
  {\bibinfo  {journal} {Z. Phys. B}\ }\textbf {\bibinfo {volume} {68}},\
  \bibinfo {pages} {217} (\bibinfo {year} {1987})}\BibitemShut {NoStop}%
\end{thebibliography}%

\end{document}